\documentclass[aps,12pt,showpacs]{revtex4}
\usepackage[english]{babel}
\usepackage{amsmath}
\usepackage{amssymb}
\usepackage{amsbsy}
\usepackage{amstext}
\usepackage{pst-node}
\usepackage[dvips]{graphicx}

\begin{document}

\title{All-sky incoherent search for periodic signals with
Explorer 2005 data}

\author{P~Astone$^1$, D~Babusci$^3$, M~Bassan$^{4,5}$,
P~Carelli$^{6,5}$, G~Cavallari$^8$, A~Chincarini$^2$, E~Coccia$^{4,5}$,
S~D'Antonio$^5$, M. Di Paolo Emilio $^{6,7}$,
V~Fafone$^{4,5}$, S~Foffa$^{10}$, G~Gemme$^2$, G~Giordano$^3$,
M~Maggiore$^{10}$, A~Marini$^3$, Y~Minenkov$^7$,
I~Modena$^{4,5}$, G~Modestino$^3$, A~Moleti$^{4,5}$, G~P~Murtas$^3$, G~V~Pallottino$^{9,1}$, R~Parodi$^2$,
G~Piano~Mortari$^{6,7}$, G~Pizzella$^{4,3}$,
L~Quintieri$^3$,  A Rocchi$^{4,5}$,
F~Ronga$^3$, F. Saint Just $^{9,1}$, R~Sturani$^{10}$, R~Terenzi$^{11,4}$, G~Torrioli$^{12,1}$,
R~Vaccarone$^2$, G~Vandoni$^8$ and M~Visco$^{11,5}$}

\address{$^1$ INFN, Sezione di Roma, Roma, Italy}
\address{$^2$ INFN, Sezione di Genova,  Genova, Italy}
\address{$^3$ INFN, Laboratori Nazionali di Frascati, Frascati, Italy}
\address{$^4$ Dip.  Fisica, $\mathrm{Universit\grave{a}}$ di Roma ``Tor
Vergata'', Roma, Italy}
\address{$^5$ INFN, Sezione di Roma Tor Vergata, Roma, Italy}
\address{$^6$ $\mathrm{Universit\grave{a}}$ dell'Aquila, Italy}
\address{$^7$ INFN, Laboratori Nazionali del Gran Sasso, Assergi,  L'Aquila,
Italy}
\address{$^8$ CERN, Geneva , Switzerland}
\address{$^9$ Dip. Fisica, $\mathrm{Universit\grave{a}}$  di Roma ``La
Sapienza'', Roma, Italy}
\address{$^{10}$ Dep. de Phys.
$\mathrm{Th\acute{e}orique,~Universit\acute{e}~de~Gen\grave{e}ve,~Gen \grave{
e}ve,~Switzerland}$ }
\address{$^{11}$ INAF, Istituto Fisica Spazio Interplanetario, Roma,  Italy}
\address{$^{12}$ CNR, Istituto di Fotonica e Nanotecnologie,  Roma,  Italy}

\begin{abstract}
The data collected during 2005 by the resonant bar Explorer
are divided into segments and incoherently summed in order to
perform an all-sky search for periodic gravitational wave signals.

The parameter space of the search spanned about $40$Hz in frequency,
over 23927 positions in the sky. Neither source orbital corrections
nor spindown parameters have been included, with the result that the
search was sensible to isolated neutron stars with a frequency drift less than
$6\cdot10^{-11}$Hz/s.

No gravitational wave candidates have been found by means of the present
analysis, which led to a best upper limit of $3.1\cdot 10^{-23}$ for the
dimensionless strain amplitude.
\end{abstract}
\maketitle

\subsection{Introduction}
The search for periodic gravitational wave signals is a stimulating challenge
for data analists because of the considerable amount of computing time
required.

For blind searches, i.e. without any {\it a priori} knowledge about the source,
a fully coherent analysis can not handle more than a few days of data because
of the steep dependence of the size of the parameter space on the
frequency resolution.

In \cite{Astone:2005fn} three data sets, each two days long, from the
Explorer 1991 run have been coherently studied by means of the 
{\cal F} statistics method \cite{Jaranowski:1998qm}
which led to and an upper limit of $1\cdot 10^{-22}$ on $h$ in
the narrow band $921.00$-$921.76$Hz.

A similar technique, applied in \cite{Abbott:2006vg}
to the most sensitive 10 hours of the the LIGO S2 run, led
to an upper limit of $6.6\cdot 10^{-23}$ for isolated
neutron stars in the band between $160$ and $728.8$Hz.

With the widening of the frequency band, due to the advent of
interferometers as well as to improvements in the readout of resonant
detectors, several incoherent and semi-coherent methods have been
conceived and employed.

In \cite{Abbott:2005pu} the Hough transform technique has been applied to
the LIGO S2 data to perform a blind search for isolated neutron stars
on a set of narrow frequency bands in the range $200$-$400$Hz, and a best
upper limit of $4.43\cdot 10^{-23}$ has been set.

In the present work,  the simple technique of adding power spectra has been
applied to the most sensitive $40$Hz band of the 2005 run of the Explorer bar
\cite{Astone:2006uf}, resulting in a further improvement on the best upper
limit on $h$, which is set to $3.1\cdot 10^{-23}$ at $920.14$Hz.

According to the results reported at the recent Amaldi7 conference,
the analysis of the LIGO S4 run \cite{Abbott:2007td} is leading to a sensible
improvement in this direction (about an order of magnitude);
remarkably, the limit set in the present work is still competitive with
the one coming from the S4 data in the same frequency band.

\subsection{The data set}
At the end of April 2005, after a short commissioning break, the resonant
antenna Explorer was again
{\it on air} and operated with the usual, remarkable duty cycle ($86\%$
from April to December 2005) and a good stability.

The data stream taken by the bar until the end of 2005
(after which the sensitivity curve has been modified)
has been divided into $25161$ segments, each
about 14 minutes long.
The most sensitive band of the Fourier transform of these segments,
namely $N_f=32178$ frequency bins in the range $885$-$925$Hz,
has been selected for the analysis.
 
\begin{figure}
   \centering
   \includegraphics[width=5in,angle=-90]{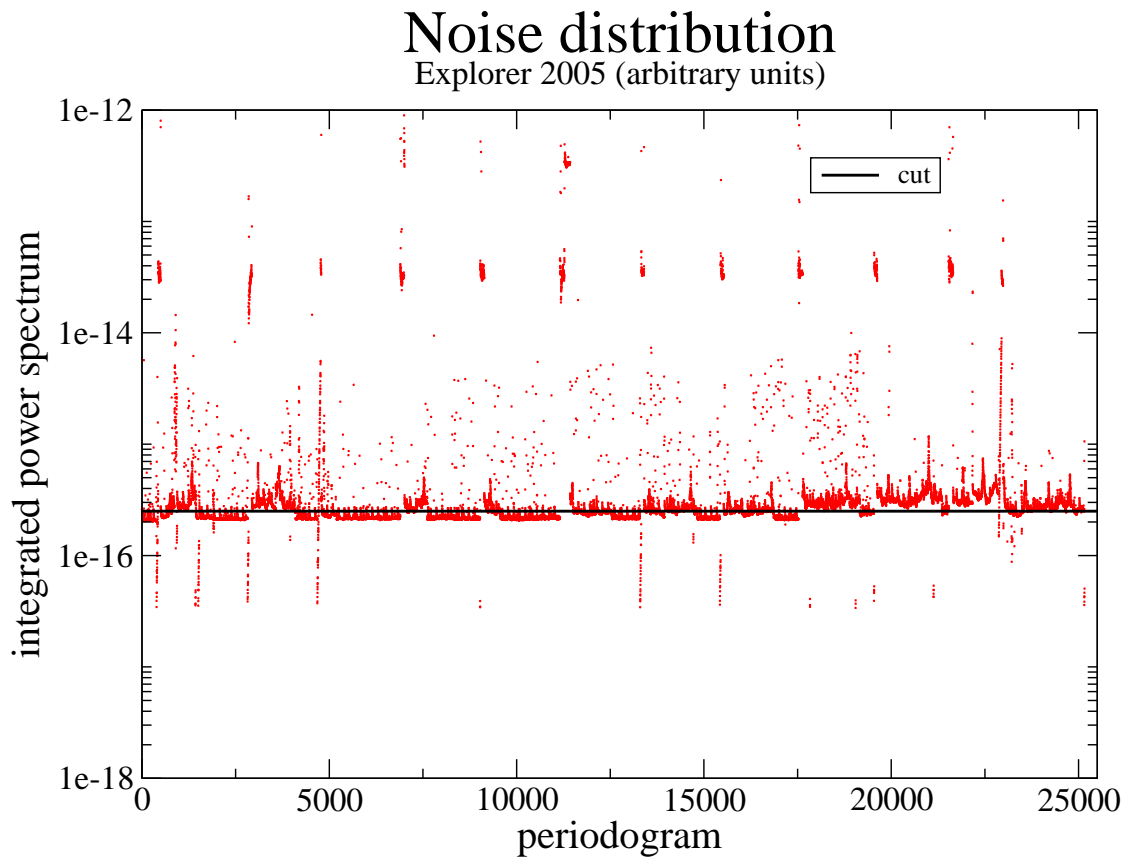} 
   \caption{The noise level of Explorer during the 2005 run.
The solid horizontal line is the cut applied to select the best spectra.}
   \label{fignoise}
\end{figure}

A noise cut has been applied to the total power contained in each 
spectrum, with the purpose of discarding the noisy ones thus creating an
homogeneous
set of spectra. This allowed us to apply for the subsequent analysis
the simple power addition method, without weighting each spectrum with the
corresponding noise level.

This selection led to the creation of a data set $\left\{S_i\right\}$
made of the $N_1=11749$ cleanest spectra (corresponding to an effective data
time of 114 days), and of a second set
containing just the best $N_2=3875$ (used to deal with the critical zones
of the spectrum, corresponding to the resonant modes of the bar, around 888Hz
and 920Hz).

\begin{figure}
   \centering
   \includegraphics[width=6in,angle=0]{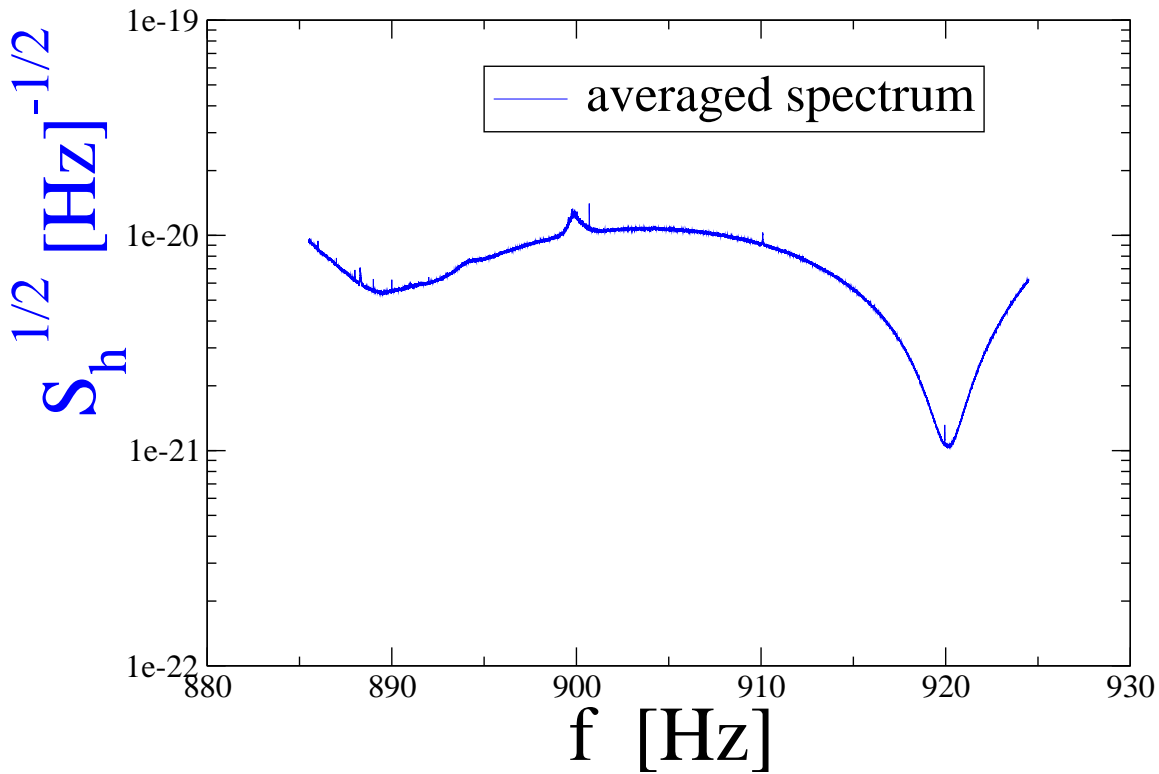} 
   \caption{The square root of the average spectrum represents the typical
Explorer sensitivity curve during the 2005 run.}
   \label{figsensi}
\end{figure}

The average sensitivity of the second data set is shown in Fig. \ref{figsensi}
(the first set is similar except around the resonant modes);
a few noisy lines may be noted, including small 1Hz harmonics on the left part.

\subsection{The analysis method}

For a given direction $\hat{r_j}$ in the sky, the selected spectra have
been deformed according to the Doppler shift formula
\begin{eqnarray}\label{doppler}
f^{\rm true}_j\simeq f_{\rm exp}\left(1-\frac{\vec{v}_{rel}
\cdot\hat{r_j}}{c}\right)\nonumber
\end{eqnarray}
and then summed and renormalized dividing by $N_1$ or by $N_2$:
\begin{eqnarray}\label{sumspe}
{\cal S}_j (f^{\rm true}_j)=\frac{1}{N_{1,2}}\sum_{i=1}^{N_{1,2}}
S_i(f^{\rm true}_j)\, .\nonumber
\end{eqnarray}

The speed of the detector relative to the Solar System Baricenter has been
computed thanks to the JPL ephemerides \cite{JPL}.
As the speed of the source have not been taken into account, the search is
sensitive to isolated neutron stars, and not to those which are part of
binary systems.

Spindown has also been neglected: given the frequency resolution of
$1.2\cdot 10^{-3}$Hz, this means being sensitive to an average frequency
drift up to $6\cdot10^{-11}$Hz/s during the observation
period.\\
The procedure has been repeated for any point of an optimized sky grid
made of $N_{\rm sky}=23927$ possible directions, thus leading to the creation
of a set $\left\{{\cal S}_j\right\}$ containing
$N_{\rm sky}$ ``deformed and summed'' spectra.

\begin{figure}
   \centering
   \includegraphics[width=5in,angle=0]{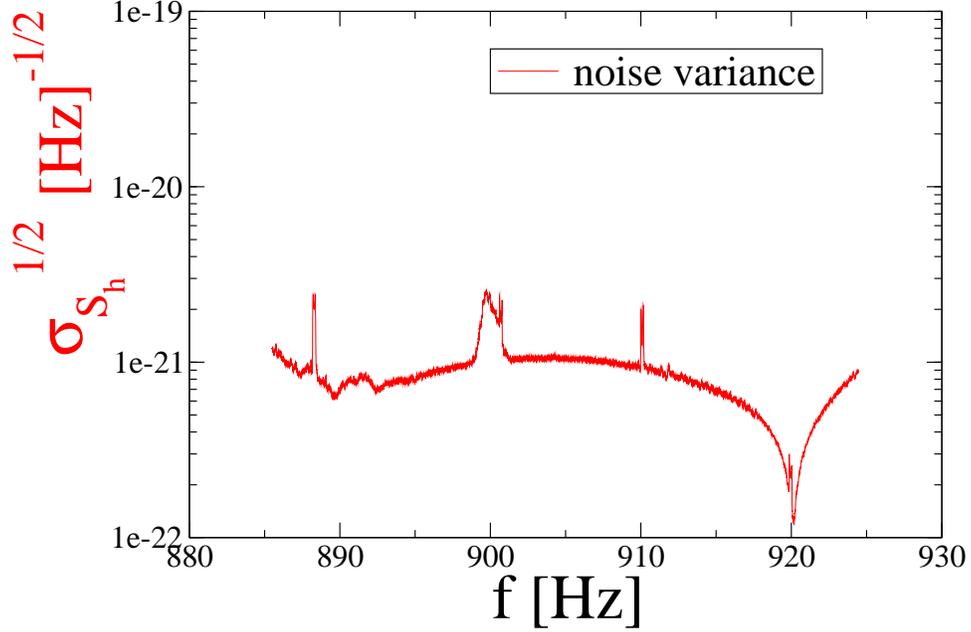} 
   \caption{Noise variance of the set of deformed and summed spectra.
To help comparison with Fig. \ref{figsensi}, the square root of $\sigma$ is
actually shown.}
   \label{figsigma}
\end{figure}

The variance of the noise is obtained calculating, for each value of the
frequency, the variance of the distribution of the $N_{\rm sky}$ deformed
spectra.
The result, whose square root is shown in Fig. \ref{figsigma}, agrees with
the general expectation, based on the central limit theorem, that
\begin{equation}
\sigma\simeq\frac{S_h}{\sqrt{N_{1,2}}}\, .
\end{equation}

As expected, the plot shows an anomalous behavior of the noise
variance in correspondence of the disturbances of the initial data set.
These anomalous zones have not been taken into account for the candidate
search, but have been included in the upper limit determination.

The detection threshold is fixed by the requirement that the false alarm rate
should be less than $1\%$. According to Poisson statistics one has to impose

\begin{eqnarray}\label{poisson}
P(0,\lambda)={\rm e}^{-\lambda}>.99\, ,\nonumber
\end{eqnarray}
where the expected number of threshold crossings in absence of signal is
\begin{eqnarray}\label{lambda}
\lambda=p\cdot N_f\cdot N_{\rm sky}\, ,\nonumber
\end{eqnarray}
being $p$ the probability of false detection in a single frequency bin
and for a single direction in the sky, and $N_f\cdot N_{\rm sky}$
the trial factor.

One thus finds the condition $p<1.3\cdot 10^{-11}$
which, assuming that the $N_{\rm sky}$ values of the shifted spectra at a
given frequency bin are gaussian distributed, translates to a $7\sigma$
threshold.

In other words, a detection is claimed if, for some value of
the frequency $f$, a given deformed spectrum ${\cal S}_j$
satisfies the condition
\begin{eqnarray}\label{detcond}
{\cal S}_j(f)-\bar{\cal S}(f)>7 \sigma(f)\, ,
\end{eqnarray}
being $\bar{\cal S}$ the average of the deformed spectra
$\left\{{\cal S}_j\right\}$ as $j$ spans over the $N_{\rm sky}$
sky grid points. 

Since $h\propto\sqrt{S}$, to translate the detection threshold
in $h$ units one has to multiply the values shown in Fig. \ref{figsigma}
by the factor
\begin{eqnarray}\label{factor}
\sqrt{\frac{7}{T}}\cdot\sqrt{\frac{15}{4}}\, ,\nonumber
\end{eqnarray}
where $T$ is the length of each data segment and the factor $4/15$
(the average angular sensitivity of the bar over the solid angle) is
introduced to compensate the fact that amplitude modulation has not been taken
into account when summing the deformed spectra.

\subsection{Results}
\subsubsection{Cadidate search}

\begin{figure}
   \centering
   \includegraphics[width=5in,angle=0]{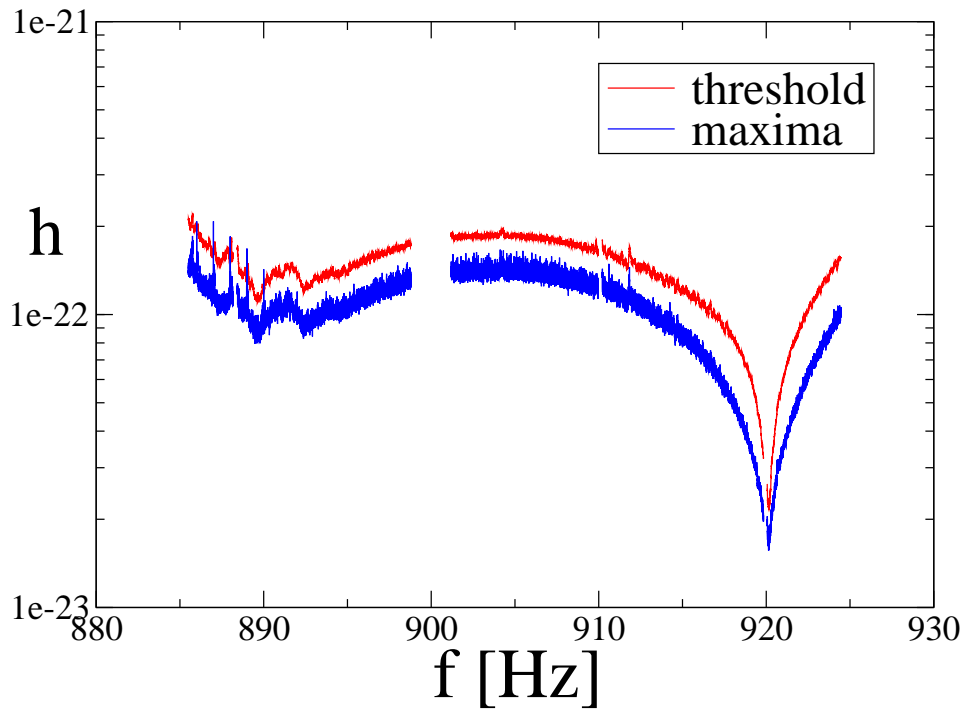} 
   \caption{Detection threshold (red, upper curve) and maximal triggers
     (blue, lower). The $1$Hz disturbances on the left have been left only
     for illustrative purpose and cannot be considered as genuine threshold
     crossings.}
   \label{figthremax}
\end{figure}

The threshold is shown in red in Figure \ref{figthremax},
while the blue line is the maximum value of $h$ found, at any given
frequency bin, among the set of the $N_{\rm sky}$ spectra according to
the following formula:
\begin{eqnarray}\label{geth}
h_{\rm max}(f)={\rm max}_j \left[\sqrt{\frac{15}{4}\frac{{\cal S}_j(f)-\bar{\cal S}(f)}{T}}\right]\, .\nonumber
\end{eqnarray}

 The anomalous regions of the spectrum have been cut out, since it is not
reasonable to assume that they follow a gaussian distribution. To illustrate
the point, the $1$Hz disturbances have been keft in Figure \ref{figthremax},
to show that they would have
produced fake candidates, if they had been included in the analysis.

With these specifications, the maxima are never above the threshold,
and thus no candidates have been found.

\subsubsection{Software injections and upper limit}

To find an upper limit on $h$, a variation of the loudest event method
\cite {Brady:2004gt} has been applied.
This method allows to determine an upper limit starting from the
loudest event present in a data stream, irrespectively of the fact that
such an event may be due to noise or to a real signal.
The idea is basically that if a strong real signal would have been present
during the data taking, it would have produced an event louder than the
loudest event actually recorded. This idea can be made quantitative by
studying the detection efficiency of the experiment, for example by performing
software injections at various SNR's.

\begin{figure}
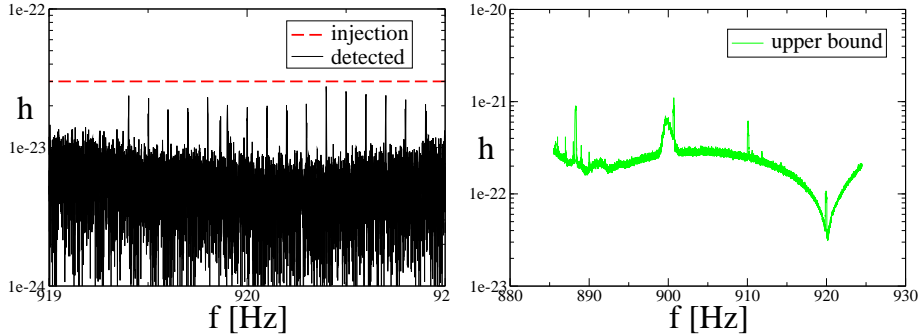

  \centering
  \includegraphics[width=6cm,angle=0]{inje.eps}
  \includegraphics[width=6cm,angle=0]{upper.eps}
   \caption{On the left, the outcome of the 16 injections,
   with initial amplitude given by the dashed red line.
   On the right, the upper limt curve.}
   \label{figinjupp}   
\end{figure}

In our case, we have a loudest event for any frequency bin, and all these
events form precisely the curve of maxima depicted in Figure \ref{figthremax}.
As a preliminary step, we had to determine our detection efficiency by
injecting fake periodic signals in the Explorer data stream and
finding out how hey did loook like after the analysis chain: the left plot of
Figure \ref{figinjupp} shows for instance the result of 16 injections
of signals with $h=3\cdot 10^{-23}$, equally spaced in frequency by
$0.1$Hz starting from $919.4$Hz, and coming from randomly chosen directions
in the sky.
Then, the upper limit at $95\%$ c.l. at has been determined for each
frequency bin as the lowest injected amplitude which had produced a
signal larger than the actual maximum at least in $95\%$ of the cases.

The right plot on Figure \ref{figinjupp} shows the result, i.e. the curve
of $h$ upper limit at $95\%$ confidence level: the minimum is
$3.1\cdot 10^{-23}$ at $920.14$ Hz.

\end{document}